\documentstyle[twoside,mpla2]{article}
\newcommand{\ave}[1]{\langle {#1} \rangle}
\newcommand{\ket}[1]{| {#1} \rangle}
\begin{document}

\runninghead{The Octet of Goldstone-Bosons in the $SU(3)$ Linear-$\sigma$-Model
$\ldots$} {The Octet of Goldstone-Bosons  in the $SU(3)$ Linear-$\sigma$-Model
$\ldots$}

\normalsize\textlineskip
\thispagestyle{empty}
\setcounter{page}{1}

\copyrightheading{}                   

\vspace*{0.88truein}

\fpage{1}
\centerline{\bf 
The Octet of Goldstone-Bosons in the ${\bf SU(3)}$ Linear-$\sigma$-Model
in the QRPA\fnm{*}\fnt{*}{
{\bf  IKDA: 97/31 and  hep-ph/9710419}}}
\vspace*{0.37truein}
\centerline{\footnotesize Z. AOUISSAT, O. BOHR, J. WAMBACH }
\vspace*{0.015truein}
\centerline{\footnotesize\it Institut f\"ur Kernphysik,
 Technische Universit\"at Darmstadt,}
\baselineskip=10pt
\centerline{\footnotesize\it Schlo{\ss}gartenstra{\ss}e 9, D-64289 
 Darmstadt, Germany}

\vspace*{0.21truein}
\abstracts{A symmetry conserving, non-perturbative treatment based on a 
variational squeezed vacuum state in conjunction with a well-defined 
class of RPA fluctuations is applied to the $SU(3)$ linear-$\sigma$-model. 
It is shown that the Goldstone theorem holds exactly both at zero and finite 
temperature. The approach represents a systematic procedure which avoids
problems of the Gaussian Functional with the symmetries.}{}{}

\vspace*{10pt}

\textlineskip   
\vspace*{12pt}
Spontaneously broken chiral symmetry and its restoration at finite temperature
and baryon chemical potential is one of the fascinating and still open
subjects in strong interaction physics. This problem can be addressed within 
effective theories with linear realization of chiral symmetry, such as 
linear-$\sigma$-models \cite{PISWIL}. Since spontaneous symmetry breaking and 
its restoration is non-perturbative by nature, non-perturbative methods for solving 
the effective field theory are called for.
However, in contrast to a perturbation expansion which is trivially 
symmetry conserving, the situation in non-perturbative approaches is  
more subtle. For instance, in a recent treatment of the three-flavor
linear-$\sigma$-model ($N_f=3$) at finite temperature \cite{ORTSCH}, 
to leading order in the $1/N_f$-expansion, symmetry
requirements could not be adequately dealt with, even though the
$1/N_f$-expansion is symmetry conserving. As a result
the Goldstone-bosons were acquiring masses due to thermal fluctuations.

In the present letter we show that such difficulties can be
avoided by using the Quasi-particle Random-Phase-Approximation (QRPA) 
\cite{RISCH}. 
While mixing orders in the coupling constants as well as in the
$1/N_f$-counting, this approach is nevertheless symmetry conserving,
as has been demonstrated in refs.~\cite{ACSW,ASW} for the two-flavor
case, $N_f=2$. The starting point for the QRPA is 
the quasi-particle basis, defined through a variational 
Hartree-Fock-Bogoliubov (HFB) mean field. The HFB theory is equivalent 
to the well-known Gaussian functional approach (GFA),
extensively used for scalar theories \cite{Giova} as well as in the 
context of gauge theories \cite{koko}. The GFA, although 
very appealing since it transcends the semi-classical approximation 
based on $1/{N}$-expansions, suffers from the fact that it is not symmetry 
conserving and hence the Goldstone theorem (GT) is violated.
This was, in fact, known since the early days of the GFA but no 
answer has been proposed since, without any guidelines, it is practically 
impossible to guess the missing contributions, necessary to preserve the 
symmetry. As will be demonstrated below, the QRPA provides a precise
framework for obtaining these contributions. \\

To study $SU(3) \times SU(3)$ chiral symmetry non-perturbatively (both
at zero and finite temperature) we start from the 
$SU(3)$-version of the linear-$\sigma$-model Lagrangian \cite{GAS69}
\begin{eqnarray}
 {\cal L}  &=&  \frac{1}{2}Tr \left(\partial_{\mu}M \partial^{\mu}M^+\right)
   \, -\, \frac{\mu^2}{2}\,Tr  M M^+ 
        \,+\, f_1 \left(Tr  M M^+ \right)^2\nonumber\\
     \ &+& \,f_2 Tr\left( MM^+MM^+\right) 
      \,+\,  g \left( det M \,+\, detM^+ \right)
       - c_0 s_0 \,-\, c_8 s_8~,
\label{eq1}
\end{eqnarray}
where $f_1, f_2 $ and $g$ are the three possible couplings, $\mu$ is the bare
mass and $M$  the matrix defined by
$M \,=\, \frac{1}{\sqrt{2}}  \lambda_j\left(s_j\,+\, ip_j\right) $. 
For the   Gell-Mann matrices $\lambda_j$ 
our conventions follow those in ref.~\cite{GAS69}. In the following, sums 
over repeated indices are assumed.
For later purposes we recall the conserved axial charge  
\begin{equation}
 Q_{5 \,a}  \,=\,\int d^3 x\, d_{abc} \,\left[s_b(x)
  \partial^{0}p_{c}(x)  - p_{b}(x)  \partial^{0} s_{c}(x) \right]~,
\label{eq5}
\end{equation}
where $d_{abc}$ is the symmetric tensor, $d_{abc} =\frac{1}{4} 
Tr\left(\lambda_a \{\lambda_b \, ,\, \lambda_c\}\right)$. \\ 
When dealing with the theory at finite temperature and 
in chemical equilibrium  the relevant quantity is the
grand canonical potential
\begin{equation}
\Omega =  \ave{H} - TS,
\end{equation}
where $T$ is the temperature, $S$ is the entropy of a gas of interacting bosons
and $\ave{H}$ the thermal average of the Hamiltonian on the grand canonical
ensemble.\\ 
To define the Hamiltonian a second-quantized formulation will be used.
Therefore the scalar ($s_i$) and pseudo-scalar ($p_i$) fields are 
represented respectively  by their creation and annihilation operators $b_{i}, b_{i}^+$ 
and $a_{i}, a_{i}^+$. In a first step, a canonical thermal 
Bogoliubov transformation \cite{GOO80} is performed by introducing a set of 
quasi-particle creation  and annihilation operators through the following 
rotation 
\begin{eqnarray}
  \alpha ^{+\,T}_{\mu}(q) &=& U^{(p)}_{\mu j}(\theta_p^T)\left[u_{j}^T(q)
  a^{+}_{j}(q) 
  -  v_{j}^T(q)a_{j}(-q)\right],
  \nonumber\\
  \beta ^{+\,T}_{\mu}(q) &=& U^{(s)}_{\mu j}(\theta_s^T)\left[
  x_{j}^T (q) b^{+}_{i}(q) -
  y_{j}^T (q) b_{j}(-q) - w_{j}^T \delta(q) \right]~.           
\label{eq7}
\end{eqnarray}
where $u_{i}^T (q)$, $v_{i}^T (q)$,  $x_{i}^T (q)$ and $y_{i}^T (q)$  are
even functions in the three-momenta, and $w_i^T$ denote c-numbers defined as
\[w_{i}^T = w_{0}^T \delta_{i0} + w_{8}^T \delta_{i8}.\]
The presence of the condensates renders the 
second Bogoliubov transformation in (\ref{eq7}) 
inhomogeneous. 
The orthogonal matrix $U$ decouples those modes which are not mass eigenstates of the 
Hamiltonian, and is given by
\begin{equation}
U^{(\Phi)}_{\rho i}(\theta_\Phi^T) = \delta_{\rho i} 
\delta_{i \underline{i}}
+
\left( \delta_{i 8} \delta_{\rho 8} + \delta_{i 0} \delta_{\rho 0} \right)
 \cos{\theta_{\Phi}^T} 
+  \left(
 \delta_{i 8} \delta_{\rho 0} - \delta_{i 0} \delta_{\rho 8} \right)
\sin{\theta_{\Phi}^T}.
\label{eq9}
\end{equation}
Here $\underline{i}$ runs from 1 to 7. 
Greek symbols will be used to identify the states which
are mass eigenstates of the quadratic part of the Hamiltonian, and Latin
symbols for those which are not. However, both letters will designate the
same set of numbers ranging from $0,..8$.
The symbol $\Phi$ denotes both the scalar and pseudo-scalar fields and will
be used whenever there is no ambiguity.
The condition,
\[u_{i}^{T\,2}(q) -  v_{i}^{T\,2}(q) \,=\,  x_{i}^{T\,2} (q)
- y_{i}^{T\,2}(q) \,=\, 1,\] 
which renders the transformation in (\ref{eq7}) canonical 
holds also here. \\
In thermal equilibrium, the distribution of maximum entropy 
is the one which  minimizes $\Omega$ 
with the entropy for a gas of different species of bosons given by
\begin{equation}
 S= k_B \sum_{\nu}\left[(1+f_{\nu})ln(1+f_{\nu})-f_{\nu}lnf_{\nu}\right]~,
\label{eq11}
\end{equation}   
where $f_{\nu}$ are the equilibrium Bose-distribution functions and the sum 
over $\nu$ includes the number of species as well as the three-momentum q.\\
The expectation value of the Hamiltonian in the grand canonical 
ensemble can be easily worked out by using the Bloch-Dominicis theorem \cite{BLDO}.
This leads to
\begin{eqnarray}
 \ave{H} &=& 
\int\frac{d^3q}{(2\pi)^3}\, \omega_{q}
\left[\left(1+2f_i^p(q)\right)\left(u_{i}^2(q) + v_{i}^2 (q)\right) 
+ \left(1+2f_i^s(q)\right)\left(x_{i}^2(q) + y_{i}^2(q)\right) \right]
\nonumber\\ 
&+& 
c_i\xi_i \,+\, \frac{\mu^2}{2} {\xi_i}^2
- {\cal G}_{ijk}\xi_i\xi_j\xi_k
- \frac{1}{3} {\cal F}_{ijkl}\xi_i\xi_j\xi_k\xi_l  
- \left( 2{\cal F}_{ijkl} \xi_k\xi_l + 3 {\cal G}_{ijk}\xi_k\right)
 I_{ij}^{(s)}
\nonumber\\
&-& 
\left( 2\overline{\cal F}_{ijkl}\xi_k\xi_l - 3 {\cal G}_{ijk}\xi_k
\right) 
I_{ij}^{(p)}
- 2 \overline{\cal F}_{ijkl}I_{ij}^{(p)} I_{kl}^{(s)} 
- {\cal F}_{ijkl}\left( I^{(p)}_{ij}I_{kl}^{(p)}
+ I_{ij}^{(s)}I_{kl}^{(s)} \right)~, \nonumber \\
\label{eq12}
\end{eqnarray}
where ${\cal F}_{ijkl}$,
$\overline{\cal F}_{ij,kl}$,  ${\cal G}_{ijk}$ are tensors 
which have also been given in \cite{CHHAY1}
\begin{eqnarray} 
{\cal F}_{ijkl} &=&\frac{1}{2}\left(3 f_1\delta_{m 0} + f_2\right)
\left[ d_{ijm}d_{klm} + d_{ilm}d_{jkm} + d_{ikm}d_{jlm}\right]~,
\nonumber\\
\overline{\cal F}_{ij,kl} &=&\frac{1}{2}\left(3 f_1\delta_{m 0} + f_2\right)
\left[ d_{ijm}d_{klm} + f_{ilm}f_{jkm} + f_{ikm}f_{jlm}\right]~,
\nonumber\\
{\cal G}_{ijk} &=&g\frac{\sqrt{2}}{3} d_{ijk} \left[1 -
\frac{3}{2}\left(\delta_{i0} + \delta_{j0} + \delta_{k0} \right) 
+ \frac{9}{2}\,\delta_{j0}\delta_{k0} \right]~.
\label{eq13} 
\end{eqnarray}
To fix the functions $u_{i}(q)$, $x_{i}(q)$, 
$\theta_s$, $\theta_p$ and $w_{i}$ we make use of the 
Ritz variational principle. Minimizing the grand  
potential $\Omega$,   
the finite-temperature BCS-equations follow straightforwardly.
The quasi-particle energies for the scalars and pseudo-scalars  
therefore read as
\begin{equation}
{\cal E}_{(\Phi)_{\rho}}^{T\,2} = \mu^2 - \Delta_{(\Phi)_{ij}}^T 
U_{(\Phi)_{i \rho}}^T
U_{(\Phi)_{j \rho}}^T~,
\label{eq15}
\end{equation}
in terms of the scalar and pseudo-scalar gap parameters
\begin{eqnarray}
 \Delta_{(s)_{ij}}^T &=& 4\left[{\cal F}_{ijkl}\xi^T_k\xi^T_l  +
\frac{3}{2}{\cal G}_{ijk}\xi^T_k  + {\cal F}_{ijkl}I_{(s)_{kl}}^T  + 
\overline{\cal F}_{ij,kl}I_{(p)_{kl}}^T\right]~, \nonumber\\
 \Delta_{(p)_{ij}}^T &=& 4 \left[ \overline{\cal F}_{ij,kl}\xi^T_k\xi^T_l -
\frac{3}{2}{\cal G}_{ijk}\xi^T_k 
+ {\cal F}_{ijkl}I_{(p)_{kl}}^T + \overline{\cal
F}_{ij,kl}I_{(s)_{kl}}^T\right]\, .
\label{eq16}
\end{eqnarray} 
The one-point functions are denoted by  $\xi_i = \ave{\sigma_i}$, while 
$I_{(\Phi)_{ij}}^T$ are thermal averages for the bilinears
of pseudo-scalar and scalar fields. At the minimum, these can be expressed
 explicitly in terms of the $U$-matrices as 
\begin{equation}
I_{(\Phi)_{ij}}^T \,=\, \int d{\bf x} \,\, \ave{ \Phi_i({\bf x}) 
\Phi_j({\bf x}) } \,=\,
U_{(\Phi)_{i \rho}}^T U_{(\Phi)_{j \rho}}^T I_{(\Phi)_{\rho}}^T~,
\label{eq14}
\end{equation}
where $I_{(\Phi)_{\rho}}^T$ are thermal tadpoles given in terms of 
the self-consistent quasi-particle energies and the
Bose factors $ f_{(\Phi)_{\rho}}^T(q)$ by
\begin{equation}
I_{(\Phi)_{\rho}}^T= \int\frac{d^3q}{(2\pi)^3} 
    \frac{1 + 2 f_{(\Phi)_{\rho}}^T(q)}{2\,{\cal E}_{(\Phi)_{\rho}}^T(q)}\, ,
\quad\quad with\quad
f_{(\Phi)_{\rho}}^T(q)= \left[
\exp{\frac{{\cal E}_{(\Phi)_{\rho}}^T(q)}{k_B T}} \,-\,1\right]^{-1}\,.
\end{equation}
It is important to notice that this self-consistency 
 is not forced by hand but rather follows directly from
the Ritz principle, as was shown  in \cite{ACSW,GOO80}.
The mixing angles are fixed through the equation 
\begin{equation}
\Delta_{(\Phi)_{ij}}^T  \left[\delta_{i0}\delta_{j0} 
    - \delta_{i8}\delta_{j8} \right] \sin{2\theta_{\Phi}}^T     =
\Delta_{(\Phi)_{ij}}^T  
    \left[\delta_{i0}\delta_{j8} +
    \delta_{i8}\delta_{j0}\right]\cos{2\theta_{\Phi}}^T~. 
\label{eq19}
\end{equation}
Finally, the minimization with respect to $w_{i}$ leads to an additional 
condition which will determine the strange- and  non-strange condensates,
related to $\xi_0 =\ave{\sigma_0}$ and $\xi_8=\ave{\sigma_8}$ via
\begin{eqnarray}
c_i \,&=&\, - \mu^2 \xi^T_i  + \frac{4}{3}{\cal F}_{ijkl}\xi^T_j\xi^T_k\xi^T_l 
 + 3 {\cal G}_{ijk}\xi^T_j\xi^T_k +
\left( 4{\cal F}_{ijkl}\xi^T_j + 3{\cal G}_{ikl}\right)
I_{(s)_{kl}}^T\nonumber\\
\,&+&\, 
\left( 4\overline{\cal F}_{ijkl}\xi^T_j - 3{\cal G}_{ikl}\right) 
I_{(p)_{kl}}^T
\quad\quad\quad\quad\quad\quad
for \quad i = 0~ or~ 8.
\label{eq20}
\end{eqnarray}
The BCS equations (\ref{eq15}, \ref{eq19}, \ref{eq20}) form twelve coupled, self-consistent 
equations.    
At zero temperature, they completely define the variational ground state 
of the theory $\ket{\Psi}$ as a squeezed state \fnm{1}\fnt{1}{ 
The squeezed vacuum is implicitly defined by the relations :  
$\alpha_{\rho}(q) \ket{\Psi} = \beta_{\rho}(q) \ket{\Psi}= 0$, for any
state $\rho$. It is obtained by successive unitary transformations 
to the trivial vacuum, such that: 
$\ket{\Psi} = U_{mx} U_{sh} U_{sq}\ket{0}$. The first two $U_{sq}$ and $U_{sh}$
are respectively a squeezing and a shift unitary transformation \cite{Blaizot}, 
while $U_{mx}$ is a mixing unitary transformation \cite{BHV}. Since these three
can be expressed as exponentials of, at most, bilinear forms in the creation
and annihilation operators, the vacuum $\ket{\Psi}$ can therefore be put in 
a more
compact \cite{BB}, but less appealing form which, in the case of a 
$SU(3)$-invariant vacuum, as treated at the end of this paper,  reads   
$\ket{\Psi} = {\cal N} \exp \left[  \sum_q 
\frac {v_j(q)}{2 u_j(q)} a_j^+ (q) a_j^+ (-q)
+ \frac {y_j(q)}{2 x_j(q)} b_j^+ (q) b_j^+ (-q) + 
\frac{w_0 (0)}{x_0} b_0(q)\delta(q)^+ \right] \ket{0}$.\\ 
 ${\cal N}$ is here  an irrelevant normalization factor.}.
In the context of functional field theory approaches  
this ground state is equivalent to the Gaussian functional 
which has been actively  studied in the context of the triviality of
$\lambda \Phi^4$ theories \cite{STEV,SAT}. According to these studies,
a spontaneously broken phase in the vacuum 
of the renormalized $O(N)$-vector model can be secured. 
This is a very interesting feature, since it is ruled out
in the $1/N$-expansion, where the vacuum of the (renormalized) 
theory lies in the same group of invariance as the Lagrangian. 
An additional complication arises, however, in case of broken continuous 
symmetries. The squeezed vacuum is not compatible with the 
Goldstone theorem which calls for the appearance of as many 
Goldstone modes as there are   broken generators.
This is clearly not the case here, as was shown in many 
instances (see \cite{ACSW,DMIT,BIRH}). 
We will see below how this can be corrected.\\ 
It is well known that the self-consistent HFB mean field 
is a 'mixing-order' approximation (see for instance \cite{ASW})
in the sense that there is no ordering in terms of the coupling 
constants nor in the number of flavors (colors or charges in 
other situations). At first sight this leads to difficulties,
as there are apparently no
guidelines for classifying the relevant missing contributions 
for preserving the symmetries. A powerful approach, used
for the first time in the context of the $SU(2)$ linear-$\sigma$-model 
in ref.~\cite{ACSW}, has proven successful in this regard. 
We will see that it is capable of handling the much more complicated 
$SU(3)$-case as well. The basic idea for restoring the symmetry 
(broken at the HFB-level) is to use the symmetry generator for 
defining a general RPA excitation operator. 
As is well known in many-body physics, RPA fluctuations, built 
on a self-consistent mean-field, restore broken symmetries. 
If the latter are continuous this is realized through the
appearance of spurious excitations of zero energy in the
RPA solutions. We briefly sketch how this can be achieved.
First one recalls the RPA equations which, for the finite temperature, read
\cite{SOM83}      
\begin{equation}
 \ave { \left[ \delta Q_{\nu} \, \, , \, \, \left[H\, , \, Q_{\nu}^+
 \right] \right]  } = m_{\nu}
 \ave{  \left[ \delta Q_{\nu}  \, , \, Q_{\nu}^+
 \right]  }~,
\label{eq21}
\end{equation}
where the expectation values are taken in the grand canonical ensemble
and $m_\nu$ denotes the excitation energy (mass) of a given state. 
The excitation operator $Q_{\nu}^+$ is chosen so as to generate 
pseudo-scalar excitations corresponding to the $\pi$, $K$, $\bar K$, 
$\eta$ and $\eta '$ mesons. The RPA equations can be derived from a 
variational principle in a restricted Fock space. A further approximation
is employed, the so-called quasi-boson approximation 
$\ave{ \left[Q_{\nu}  ,  Q_{\mu}^+ \right]  }= \delta_{\nu\mu}$, which allows 
for a linearization  without destroying
the symmetry conserving properties of the thermal RPA.  
The RPA equations (\ref{eq21})  will generate spurious solutions 
if the operator $Q_{\nu}^+$  
contains the symmetry generator specified by $Q_{5 a}$ in (\ref{eq5}).
In the chiral
limit ($c_0=c_8=0$) the {\it lhs} of (\ref{eq21}) will then be zero 
since the Hamiltonian commutes with the symmetry operator. If the norm on 
the {\it rhs} is well behaved then this forces $m_{\nu}$ to be zero.
Normalizability is 
guaranteed, however, by the use of the HFB quasi-particle basis. Hence 
$m_{\nu}$ vanishes in the chiral limit 
at all temperatures as long as the symmetry is spontaneously broken.
In the following this will be shown explicitly. \\
To proceed  
we take $Q_{\nu}^+$ as 
\begin{eqnarray}
 Q_{\nu}^+ &=& d_{\nu\mu\rho}  \left[
 X_{\mu\rho}^T\,\alpha^{+}_{\rho}(0)  -  Y_{\mu\rho}^T\,\alpha_{\rho}(0)
 \right]\nonumber\\
&+&  
\sum_{q} d_{\nu\mu\rho}
  \left[ U_{\mu\rho}^T(q) \beta^+_{\mu}(q) \alpha^{+}_{\rho}(-q)
 -  V_{\mu\rho}^T(q)\beta_{\mu}(q) \alpha_{\rho}(-q) \right]
\nonumber\\
&+&
\sum_{q} d_{\nu\mu\rho}
  \left[ W_{\mu\rho}^T(q)  \beta^+_{\mu}(q) \alpha_{\rho}(q)
  -\  Z_{\mu\rho}^T(q)  \beta_{\mu}(q) \alpha^+_{\rho}(q) \right]~,
\label{eq22}
\end{eqnarray}
where $U$, $V$, $W$, $X$, $Y$ and $Z$ are general amplitudes 
which will be fixed once the RPA eigenvalue problem solved. 
Expressing $Q_{5 a}$~ (\ref{eq5}) in second-quantized form  it is 
evident that $Q_{\nu}^+$ contains the same excitations
as $Q_{5 a}$ \fnm{2}\fnt{2}{
Starting from $Q_{5a}$ in (\ref{eq5}), one first expresses the fields in 
second-quantized form using $a^+, \ a, \ b^+, \ b$. Then one 
inserts the inverse Bogoliubov-transformation (\ref{eq7}). This leads to an
expression for $Q_{5a}$ similar to the one  for $Q_{\nu}^+$ in 
(\ref{eq22}). However, in this case the amplitudes $U$, $V$, $W$, $X$, $Y$ and 
$Z$ will have well-defined forms while in the general case of $Q_{\nu}^+$ 
they remain to be fixed by the eigenvalue problem (\ref{eq21}).}.
There is an interesting aspect which is apparent 
from the ansatz (\ref{eq22}) of the excitation operator. The latter,
in addition to the RPA bilinear forms which as usual induce 
quasi-particle scattering
processes, contains linear forms which indicate that the RPA scattering 
equation is in fact coupled to a Dyson equation.
Note also the (crucial) presence of the 
tensor $d_{\nu\mu\rho}$ which will
act as a projector on  the relevant RPA scattering-states, necessary for
fulfilling the GT. It is clear that one has a 
non-trivial criterion for selecting the relevant fluctuations which are
difficult to guess without having a systematic approach.  
\\ 
For the purpose at hand, the RPA equations can be condensed as 
\begin{equation}
\sum_{\sigma\eta} M^T_{\mu\rho,\,\sigma\eta}\,d_{\nu\sigma\eta}
\,\Pi_{\sigma\eta}^T \,=\, 0~,
\label{eq23}
\end{equation}
where $\Pi_{\sigma\eta}^T$ is some eigenvector which does not need 
to be specified any further. The other quantities are
defined as follows  
\begin{eqnarray}
M_{\mu\rho,\,\sigma\eta}^T  &=& \delta_{\mu\sigma}
\delta_{\rho\eta}
+ U_{(p)_{i\rho}}^T U_{(s)_{k\mu}}^T
 \left[8\overline{\cal F}_{ij,kl}
 -  \frac{R^T_{\tau ik}R^T_{\tau jl}}
 {m_{\nu}^{T\,2} - {\cal E}_{p_{\tau}}^{T\,2} } 
\right]U_{(p)_{j\eta}}^TU_{(s)_{l\sigma}}^T\,
 \Sigma^T_{s_{\sigma} p_{\eta}} (m_{\nu}^2)~, 
\nonumber\\
R^T_{mik}&=& 6 {\cal G}_{mik} \,-\,8\overline{\cal F}_{mi,kn}\xi^T_n~,
\label{eq24}
\end{eqnarray}
where $\Sigma_{s_{\rho} p_{\eta}} (m_{\nu}^2)$ are scalar 
pseudo-scalar thermal RPA loops given by
\begin{eqnarray}
\Sigma^T_{s_{\sigma} p_{\eta}}(m_{\nu}^{T\,2})&=&
\int \frac{d^3{\vec q}}{(2\pi)^3} \left[
\frac{{\cal E}^T_{p_{\eta}}(q)+{\cal E}^T_{s_{\sigma}}(q)}
{2 {\cal E}^T_{p_{\eta}}(q){\cal E}^T_{s_{\sigma}}(q)}
\frac{1+ f^T_{p_{\eta}}(q)+f^T_{s_{\sigma}}(q)}{
m_{\nu}^{T\,2}  -({\cal E}^T_{p_{\eta}}(q)+{\cal E}^T_{s_{\sigma}}(q))^2}\right.
\nonumber\\
\,&+&\, \left.
\frac{{\cal E}^T_{p_{\eta}}(q)-{\cal E}^T_{s_{\sigma}}(q)}
{2 {\cal E}^T_{p_{\eta}}(q){\cal E}^T_{s_{\sigma}}(q)}
\frac{f^T_{s_{\sigma}}(q)-f^T_{p_{\eta}}(q)}{ 
m_{\nu}^{T\,2} -({\cal E}^T_{p_{\eta}}(q)-{\cal E}^T_{s_{\sigma}}(q))^2 }
\right]~.
\label{eq244}
\end{eqnarray}
These RPA equations, build on the
self-consistent HFB basis, are valid  for the whole nonet of pseudo-scalars.
In the chiral limit, the octet of pseudo-scalars ought to come at zero 
RPA frequency while the singlet is not a Goldstone mode 
as long as the anomaly term is present in the Lagrangian $(g\neq 0)$.
This is indeed reflected in the RPA equations since the ninth axial charge
does not commute with the Hamiltonian as long as $g \neq 0$.\\ 
To verify that the symmetry is recovered at the RPA level a 
decisive test is to see whether the octet of pseudo-scalars has
zero mass when the symmetry of the vacuum is
broken from the $SU(3)\times SU(3)$ group down to $SU(3)$ which happens 
when the 
condensate $\ave{\sigma_8}$ vanishes. The $c_8$ explicit breaking is 
consequently put to zero ($c_8=0$). 
Close inspection shows that in this situation, as in the perturbative case,
the mixing angles $\theta_s$ 
and $\theta_p$ vanish \fnm{3}\fnt{3}{Explicitly  one has  
$ {\cal F}_{0800}=\overline{\cal F}_{0800}={\cal G}_{080}=0$.
Furthermore ${\cal F}_{08kl}I_{{(\Phi)}_{kl}}$ and 
$\overline{\cal F}_{08,kl}I_{{(\Phi)}_{kl}}$ are both proportional to 
$d_{8kl}I_{{(\Phi)}_{kl}}$. On the other hand the tensor $d_{8kl}$ enforces 
$k=l\neq 0$. Assuming octet degeneracy ( $I_{{(p)}_{kk}}=I_{\pi},~ 
I_{{(s)}_{kk}}=I_{\sigma}$ for $k=1,8$) and the fact that
$\sum_{k=1}^8d_{8kk}=0 $, $d_{800}=0$, one finally has   
a self-consistent solution  with $\Delta_{ij}^{(\Phi)} \delta_{i0}\delta_{j8} =0$ 
and hence $\theta_{\Phi}=0$.
This solution is consistent with the octet degeneracy of the scalars
and pseudo-scalars and is reminiscent of the perturbative case.}.
Therefore, at the HFB level, eqs. (\ref{eq15}) generate  mass
eigenstates of the Hamiltonian in this limit.  
Accordingly, we will not distinguish any longer between states with Greek or
Latin labelling. Finally, from the HFB equations, 
one can infer the following solution which is consistent with the octet degeneracy:
\begin{eqnarray}
{\cal E}_{(p)_k}^T&=&{\cal E}_{\pi}^T,\quad\quad\quad\quad\quad\,\,
  {\cal E}_{(s)_k}^T={\cal E}_{\sigma}^T~, 
\quad\quad\quad\quad\quad\quad
 for\quad 
 k={1,..8}~,
 \nonumber\\
 {\cal E}_{(p)_0}^T&=&  {\cal E}_{p_o}^T \neq {\cal E}_{\pi}^T,\quad\quad\quad
  {\cal E}_{(s)_0}^T={\cal E}_{s_0}^T \neq {\cal E}_{\sigma}^T~.   
\label{eq27}
\end{eqnarray}
This solution is, of course, compatible with a finite  
condensate $\ave{\sigma_0}$ and an overall mass splitting between the
scalars and pseudo-scalars.  
It is obtained for a given  choice of the parameters in the Lagrangian.
Assuming this solution, the bare mass $\mu$ can be fixed
accordingly.
 Injecting the relations (\ref{eq27}) into the RPA equations 
(\ref{eq23}-\ref{eq24}) one can derive the RPA eigenvalues.
After  lengthy but straightforward algebra which consists 
essentially in solving the characteristic equation of the RPA
eigenvalue problem, one can read off the RPA frequencies. 
Denoting the masses of the octet of Goldstone bosons  generically 
by $m_{\pi}$ one finally has
\begin{equation}              
m_{\pi}^{T\,\,2}  = {\cal E}_{\pi}^{T\,\,2} + \frac{{\cal N}^T (m_{\pi}^{T\,2})}
{\Xi^T (m_{\pi}^{T\,2})}~,
\label{eq29}
\end{equation}
with the following definitions
\begin{eqnarray}
\Xi^T (m_{\pi}^{T\,\,2})  &=& 1 - \frac{8}{3}f_2 
\left[2 \overline{\Sigma}^T_{\sigma
\pi}(m_{\pi}^{T\,\,2}) 
- \overline{\Sigma}^T_{s_0  \pi}(m_{\pi}^{T\,\,2})
- \overline{\Sigma}^T_{\sigma  p_0}(m_{\pi}^{T\,\,2})\right] \nonumber\\
&-&\, 32 f_2^2 
\overline{\Sigma}^T_{\sigma  \pi}(m_{\pi}^{T\,\,2}) \left[
\overline{\Sigma}^T_{s_0  \pi}(m_{\pi}^{T\,\,2}) + 
\overline{\Sigma}^T_{\sigma  p_0}(m_{\pi}^{T\,\,2})
\right]~,\nonumber\\
\nonumber\\
{\cal N}^T(m_{\pi}^{T\,\,2})&=& 
\frac{5u^2}{3}
\overline{\Sigma}^T_{\sigma  \pi}(m_{\pi}^{T\,\,2})
\nonumber \\
&+& \frac{2(u+w)^2}{3}
 \overline{\Sigma}^T_{s_0  \pi}(m_{\pi}^{T\,\,2}) 
+ \frac{2(u+v)^2}{3}
\overline{\Sigma}^T_{\sigma  p_0}(m_{\pi}^{T\,\,2}) 
\nonumber\\
&+&
\, 8f_2\left[\frac{5}{9}w^2- (u+w)^2\right]
\overline{\Sigma}^T_{\sigma  \pi}(m_{\pi}^{T\,\,2})
\overline{\Sigma}^T_{s_0  \pi}(m_{\pi}^{T\,\,2})
\nonumber\\
&+&
\, 8f_2\left[\frac{5}{9}v^2-(u+v)^2\right]
\overline{\Sigma}^T_{\sigma  \pi}(m_{\pi}^{T\,\,2})
\overline{\Sigma}^T_{\sigma  p_0}(m_{\pi}^{T\,\,2})
\nonumber\\
&+&\, \frac{16}{9}f_2 \left(v-w\right)^2
\overline{\Sigma}^T_{s_0  \pi}(m_{\pi}^{T\,\,2})
\overline{\Sigma}^t_{\sigma  p_0}(m_{\pi}^{T\,\,2})
\nonumber\\
&-&
\, \frac{64f_2^2}{3}(v-w)^2
\overline{\Sigma}^T_{\sigma  \pi}(m_{\pi}^{T\,\,2})
\overline{\Sigma}^T_{\sigma  p_0}(m_{\pi}^{T\,\,2})
\overline{\Sigma}^T_{s_0  \pi}(m_{\pi}^{T\,\,2})~,
\nonumber\\     
\nonumber\\
u = 2\sqrt{2}g&-&4\sqrt{\frac{2}{3}}f_2\xi^T_0,\quad\quad
v = -3\sqrt{2}g,\quad\quad
w = -3\sqrt{2}g -4\sqrt{6}f_1\xi^T_0~,\nonumber\\
\overline{\Sigma}^T_{s_{\rho} p_{\eta}} (m_{\pi}^{T\,\,2})\,&=&\,
\Sigma^T_{s_{\rho} p_{\eta}}(m_{\pi}^{T\,\,2})
\left[1\,+\,8f_1 \Sigma^T_{s_{\rho} p_{\eta}} (m_{\pi}^{T\,\,2})\right]^{-1}\,
.
\label{eq291}
\end{eqnarray}
As a last step, we show that this expression supports a 
zero-energy solution. 
We briefly sketch the proof here. Further details will be deferred to
\cite{ABW}.\\
As long as the symmetry is spontaneously broken, equations 
(\ref{eq15}) for ${\cal E}^T_{\pi}$, ${\cal E}^T_{p_0}$, 
${\cal E}^T_{\sigma}$, ${\cal E}^T_{s_o}$
and (\ref{eq20}) for $\xi^T_0$ allow for a solution with finite $\xi^T_0$.
Assuming this, one can substitute the bare mass $\mu$ in (\ref{eq15}) by
its value from (\ref{eq20}). When taking into account the relations in
(\ref{eq27}) one can  express the quasi-particle energies as a function of 
the thermal RPA loops with zero argument:~ 
$\Sigma^T_{\sigma \pi}(0)$, $\Sigma^T_{\sigma p_0}(0)$ and 
$\Sigma^T_{s_0 \pi}(0)$. 
These uniquely specify the solution since the remaining possible RPA loop  
$\Sigma^T_{s_0 p_0}(0)$ can be expressed as a function of these three.
Using the identity for regularized integrals \cite{ACSW}
\begin{equation}
I_{(p)_j}^T - I_{(s)_i}^T=  
\left({\cal E}_{p_j}^{T\,\,2}  - {\cal E}_{s_i}^{T\,\,2}  \right)
\Sigma^T_{s_{i}  p_{j}} (0)~,
\label{eq31}
\end{equation}
which holds for zero as well as finite temperature, one can verify that 
the expression for the quasi-particle energies is exactly given by
\begin{equation}
{\cal E}_{\pi}^{T\,\,2} =  - \frac{{\cal N}^T (0)}{\Xi^T (0)} -
\frac{c_0}{\xi^T_0} \, .
\label{eq32}
\end{equation}
This clearly indicates that in the chiral limit ($c_0 =0$) and for a broken
phase ($ \xi^T_0 \neq 0$) a spurious solution in 
eq. (\ref{eq29}) with $m_{\pi}^T= 0$ exists and remains as such, 
protected  
by the symmetry, until the transition temperature $T_c$ is reached. 
Once the transition to the Wigner-Weyl phase occurs, the spurious solution 
in the RPA disappears
due to the fact that the corresponding eigenvector is not
normalizable \fnm{4}\fnt{4}{Some of the elements of the diagonal matrix-norm
 given by the ${\it rhs}$ of
eq.~(\ref{eq21}) are proportional to the differences 
$f_{s_{\rho}}(q)-f_{p_{\eta}}(q)$. These vanish in the Wigner-Weyl phase 
due to the degeneracy of scalar and pseudo-scalar modes. In this situation
the eigenvalue problem in (\ref{eq21}) does not  have a solution any longer 
(see ref. \cite{ACSW}).}.
Eqs. (\ref{eq29}) and (\ref{eq32}) are the mathematical 
reflection of the GT. They form in fact the Ward identity
which links the Goldstone two-point function to the 
condensate \fnm{5}\fnt{5}{Eq. (\ref{eq29}) 
is the pole equation which leads to the pion mass. 
The residue at the pole which receives contributions from
the RPA fluctuations, is absorbed in the definition of $c_0$. More details  
on Ward identities in the linear-sigma-model could be found in \cite{blee} .}
\[
-D^{-1}_{\pi}(0) = \frac{-c_0}{\ave{\sigma_0}}~.
\]
One can appreciate the non-triviality of the present result which suggests 
that, for a one-point function (condensate) evaluated at the HFB level,
the corresponding Goldstone mode has to be an excitation in the spectrum
of the RPA.\\    
As mentioned earlier, this result is valid for
the octet of Goldstone modes while the singlet pseudo-scalar 
remains massive since the corresponding symmetry generator 
does not commute with the Hamiltonian, because of the anomaly. Further details 
as well as a quantitative study of the phase transition will be presented 
elsewhere \cite{ABW}.

In summary, we have discussed the $SU(3)$ linear-$\sigma$-model 
at finite temperature within the HFB-RPA approach.
The Goldstone theorem has been shown to hold both at zero and finite
temperature below the phase transition. Above the transition the 
spurious solution of the RPA equations are no longer normalizable, therefore 
no asymptotic zero-energy solution is present in the RPA spectrum.  
An explicit proof of the Goldstone theorem has been given in the case 
of the vacuum being an invariant of the $SU(3)$ symmetry group. 
It should be appreciated that the present approach is transcending
the usual expansion in $1/N_f$. 
It is in fact order mixing but nevertheless symmetry conserving.
It shares with the Gaussian functional approach its advantages 
and corrects for its shortcomings when it comes to fulfilling the symmetry
requirements. 
We have also demonstrated that this approach can handle successfully not only 
models with $O(N)$  symmetry  \cite{ACSW} but also those 
with $SU(N)$ symmetry.
\\
\\

\underline{Acknowledgements:}
We would like to thank G. Chanfray and P. Schuck for their interest in this
work. Z. Aouissat acknowledges  financial support from GSI-Darmstadt.

\end{document}